# Ferroelectric tuning of superconductivity and band topology in a two-dimensional heterobilayer

Jianyong Chen[1,2], Ping Cui[1,*] and Zhenyu Zhang[1,*]

[1]*International Center for Quantum Design of Functional Materials (ICQD), Hefei National Laboratory for Physical Sciences at Microscale (HFNL), University of Science and Technology of China, Hefei, Anhui 230026, China*
[2]*College of science, Guilin University of Aerospace Technology, Guilin 541004, China*

[*]Emails: cuipg@ustc.edu.cn; zhangzy@ustc.edu.cn

**ABSTRACT**

Realization of tunable superconductivity with concomitant nontrivial band topology is conceptually intriguing and highly desirable for superconducting devices and topological quantum computation. Based on first-principles calculations, here we present the first prediction of simultaneously tunable superconducting transition temperature ($T_c$) and band topology in a superconducting $IrTe_2$ overlayer on a ferroelectric $In_2Se_3$ monolayer. We first demonstrate that the $T_c$ is substantially enhanced from that of $IrTe_2$ nanoflakes ($T_c$ ~3 K) due to significant charge repartitioning around the Fermi level. More importantly, the $T_c$ is shown to sensitively depend on the $In_2Se_3$ polarization, with the higher $T_c$ of ~(8-10) K attributed to enhanced interlayer electron-phonon coupling when the polarization is downward. The band topology is also switched from trivial to nontrivial as the polarization is reversed from upward to downward. These findings provide physically realistic platforms for simultaneously tuning superconductivity and band topology in two-dimensional heterobilayers and related heterostructures using a reversible and nonvolatile approach.

Achieving tunable superconductivity is a prerequisite for fabrication of reprogrammable superconducting circuits and utilization of magnetic flux quanta [1,2]. Ferroelectric effects, as characterized by switchable polarization of a ferroelectric material upon application of a voltage pulse, can be exploited as a nonvolatile and reversible approach to modulate the superconducting transition temperature ($T_c$). Such ferroelectric tuning of superconductivity has been achieved in heterostructures of conventional superconducting and ferroelectric films. Representative examples include significant $T_c$ modulations in $Pb(Zr_xTi_{1-x})O_3/GdBa_2Cu_3O_{7-x}$ [3] and $BiFeO_3/YBa_2Cu_3O_{7-x}$ [4], and a complete switching of a superconducting transition in Nb-doped $SrTiO_3$ with $Pb(Zr,Ti)O_3$ as the ferroelectric overlayer [5].

Conventional superconductor/ferroelectric heterostructures are typically hampered by non-uniform film thickness, structural imperfections, random interfacial charge traps, and relatively broad superconducting transitions due to the two-dimensional (2D) nature of the interfacial superconductivity [6,7]. Since the ferroelectric effects are more pronounced near the interfaces [8], atomically thin superconductors are expected to be more substantially tunable. In this regard, the concurrent discoveries of 2D ferroelectric materials [9-14] and 2D superconducting materials [15-18] offer unprecedented opportunities for exploration of ferroelectrically tuned superconductivity and related devices, especially given the atomically sharp interfacial qualities of such van der Waals heterostructures [19,20]. These layered systems, in turn, may also enable realizations of other emergent physical properties and functionalities, including notably topological superconductivity for fault-tolerant topological quantum computing.

In this Letter, we use first-principles approaches to demonstrate a simultaneous tuning of $T_c$ and band topology by ferroelectric polarization, using a heterobilayer of IrTe$_2$/In$_2$Se$_3$ as the first known example based on 2D systems. As constituent building blocks, In$_2$Se$_3$ monolayers have been fabricated [11,21,22], with their ferroelectric switching experimentally demonstrated [11,21], while IrTe$_2$ nanoflakes down to the monolayer limit have also been successfully prepared [15,16,23], with the observed $T_c$ ~3 K for the nanoflakes [15,16]. We first show that the $T_c$'s of the heterobilayers are substantially enhanced compared to the IrTe$_2$ nanoflakes, which can be attributed to significant charge repartitioning around the Fermi level, as characterized by the enhanced/emergent nesting behavior of the electrons in the heterobilayers. More intriguingly, the $T_c$ sensitively depends on the In$_2$Se$_3$ polarization, with the higher $T_c$ of (7.70-9.58) K caused by enhanced interlayer electron-phonon coupling (EPC) for the downward polarization. Furthermore, the band topology is switchable from trivial to nontrivial as the polarization is reversed from upward to downward, resulting from the cooperative effects of proper band alignments and inherently strong spin-orbit coupling (SOC). These findings provide physically realistic platforms for simultaneously tuning superconductivity and band topology in 2D heterostructures using a reversible and nonvolatile approach, and open a new avenue for nanoscale superconducting device applications.

The IrTe$_2$ and In$_2$Se$_3$ monolayers share similar triangular atomic arrangement within each atomic layer, with the optimized lattice constants of 3.84 and 4.10 Å, respectively. An IrTe$_2$/In$_2$Se$_3$ heterobilayer can adopt a (1×1) matching relationship, and the lattice mismatch of ~6% is tolerable given the relatively weak interlayer coupling. Here, by simulating the case of In$_2$Se$_3$ as the substrate, we choose 4.10 Å as the lattice constant of the heterobilayer. To identify the energetically stable structures of the IrTe$_2$/In$_2$Se$_3$ heterobilayers, three high-symmetry stacking configurations (Fig. S1) have been considered for each polarization. The most stable stacking for either polarization is identical, with the inner Te atom (Te$_i$) sitting on the center of the top buckled hexagonal lattice of In$_2$Se$_3$ [Figs. 1(a) and 1(c)]. Similar polarization-insensitive interfacial structural registry has also been observed in recent studies [24,25]. The interlayer distance increases from 2.54 to 2.67 Å when the polarization is switched from downward to upward, with other structural parameters staying intact (Fig. S2). The formation energies are 0.91 and 0.81 eV for the downward and upward polarizations, respectively, indicating that both heterobilayers are energetically stable (see computational methods and details in the Supplemental Material (SM) [26]). The corresponding phonon spectra are further calculated [Figs. 1(b) and 1(d)], indicating that both systems are dynamically stable. It is noted that the negligibly small imaginary frequencies near the Γ point for the upward case are an artifact of the simulations or lattice instabilities related to long wave undulations [40-42]. Moverover, their thermodynamic stabilities are confirmed by *ab initio* molecular dynamics simulations at 300 K (Fig. S3).

Before investigating the ferroelectric tunning effects on the IrTe$_2$/In$_2$Se$_3$ heterobilayers, we first calculate the EPC and $T_c$ of IrTe$_2$ nanoflakes to benchmark our methods and facilitate further comparison with the heterobilayers. Since the IrTe$_2$ nanoflakes are thick enough (e.g., 130~160 monolayers [15]), a bulk model is appropriate to mimic these nanoflakes. The total EPC strength ($\lambda$) is calculated to be 0.76, with the three lowest-lying acoustic branches contributing 66% and the two lowest-lying optical branches also contributing substantially (Fig. S4). Using the McMillian-Allen-Dynes parametrized Eliashberg equation [37,38], we can estimate the $T_c$ to be 2.99~4.48 K, with the Coulomb pseudopotential $\mu^*$ taken as 0.15~0.10. These benchmark evaluations agree well with the experimentally observed $T_c$ ~3.0 K for IrTe$_2$ nanoflakes [15,16], and the phonon spectra overall reproduce well the experimental Raman spectra [16,43,44]. Therefore, the adopted standard approaches within density functional theory (DFT) coupled with the isotropic superconducting picture are able to properly describe the IrTe$_2$ systems. Indeed, previous DFT plus dynamical mean field theory calculations has shown negligible electron correlation effects in bulk IrTe$_2$ [45].

Next, we explore the ferroelectric-field-tuned EPC and $T_c$ of the IrTe$_2$/In$_2$Se$_3$ heterobilayers. The convergences of both the phonon dispersions and EPC constants have been carefully examined (Fig. S5 and Tables S2-S4), ensuring that the calculations are reliable. The phonon spectra, Eliashberg function $\alpha^2F(\omega)$, and $\lambda$ of the IrTe$_2$/In$_2$Se$_3$ heterobilayers with opposite polarizations are given in Figs. 1(b) and 1(d). It is noted that the density of states at the Fermi level $N(\varepsilon_F)$ of the heterobilayer with either polarization has been boosted significantly by more than 60% from that of bulk IrTe$_2$ (Table 1), indicating significant charge repartitioning around the Fermi level. The underlying reason is tied to the enhanced/emergent nesting behavior of the electrons around the Fermi level, as elaborated in more details later. Such enhancements, in turn, play a vital role in enhancing the overall $\lambda$ [46]. More intriguingly, the $T_c$ of the heterobilayer for the downward polarization is higher than that for the upward case (Table 1). The calculated partial charges for both polarizations around the Fermi level show that the charge more noticeably pervades in the interlayer space for the downward case [Fig. 2(a)]; accordingly, the interlayer vibrational modes are more likely to influence these states, contribute to the $\lambda$, and enhance $T_c$. The emergence of the increased interlayer states can also be understood by inspecting the projected density of states of the inner-layer Se atoms (Se$_i$) before and after forming the heterobilayer (Fig. S6). At a branch specific level, the eight low-frequency phonon branches have been softened for the downward polarization as compared with the upward case [Fig. 1(b)]. The EPC strengths contributed by these eight branches are 0.93 and 0.74 for downward and upward polarizations, respectively, while the rest sixteen branches collectively contribute an equal EPC strength of 0.30 for both cases.

Given the above analyses, we focus on the eight low-frequency branches to explore the physical origins of the different $\lambda$'s in the two polarizations, with attention to the enhanced/emergent Fermi nesting effects. The Fermi surface of the upward-polarized structure is composed of a large flower-like hole pocket and a smaller hexagonal hole pocket, providing strong Fermi nesting of the electrons [see $\mathbf{q}_1$ and $\mathbf{q}_2$ in Fig. 2(c)]. For the downward case, two additional hexagonal hole pockets centered at the $\Gamma$ point are introduced with the nesting vectors $\mathbf{q}_3$ and $\mathbf{q}_4$. Furthermore, the branch ($v$)- and momentum ($\mathbf{q}$)-resolved $\lambda_{\mathbf{q}v}$ are indicated by the sizes of the red circles in Figs. 1(b) and 1(d). The most prominent $\lambda_{\mathbf{q}v}$'s lie between (1/4-3/4) regions along the $\Gamma$K and $\Gamma$M directions, corresponding to the well-defined nesting vectors $\mathbf{q}_1$ and $\mathbf{q}_2$ along $\Gamma$K and EPC-involved phonon momenta along $\Gamma$M. In particular, the five phonon modes ($\alpha_1$, $\alpha_2$, $\beta_1$, $\beta_2$, and $\gamma$) depicted in Fig. 2(b) significantly enhance the EPC when the polarization direction changes. Here, $\alpha_1$, $\alpha_2$, $\beta_1$, and $\beta_2$ are mainly associated with the interlayer out-of-plane vibrations (mixed with the IrTe$_2$ intralayer vibrations in $\beta_1$ and $\beta_2$), while $\gamma$ is dominated by the interlayer in-plane vibrations. In addition, our branch-resolved $\lambda$ calculations (Table S5) show that the most striking enhancement in $\lambda$ originates from branch 1, from 0.112 (upward) to 0.224 (downward), accounting for 50% of the total EPC increase [Fig. 1(e)]. Besides, the total contribution from branches 2-8 is also enhanced (Fig. S6), collectively resulting in a moderate increase in $\lambda$. The latter can be attributed to a slight increasing of $N(\varepsilon_F)$ associated with the additional Fermi nesting of $\mathbf{q}_3$ and $\mathbf{q}_4$; in contrast, the abnormally pronounced enhancement of the $\lambda$ for branch 1 cannot be explained by the tiny increasing of $N(\varepsilon_F)$ and Fermi nesting.

To decipher the underlying mechanism of the branch-1 enhanced $\lambda$, we analyze the $\lambda_{\mathbf{q}v}$ by investigating the EPC matrix. Usually, the EPC matrix elements can be inferred from the shifts of the energy bands in a frozen phonon calculation [36]. The band structures before and after the distortion by the selective interlayer out-of-plane vibration $\alpha_1$ of branch 1 at the 1/3$\Gamma$M point are calculated using a commensurate 6×1×1 supercell [Fig. 2(d)]. With the atom displacement of 0.05 Å, the band degeneracies at the A point are lifted and band splittings are induced. The $\lambda_{\mathbf{q}v}$ is approximately proportional to the square of the band splittings in the frozen phonon calculation [47]. From Fig. 2(d), we obtain that the ratio of the average splitting energies between the downward and upward polarizations is 1.5, leading to an enhancement of the $\lambda_{\mathbf{q}v}$ by a factor of ~2.3. This enhancement is consistent with the result of our density functional perturbation theory (DFPT) calculations. Therefore, when the polarization is switched from upward to downward, the pronounced enhancement of the EPC from branch 1 originates from the increased EPC matrix induced by the interlayer vibrations, i.e., enhanced interlayer electron-phonon coupling.

At this point, it is worthwhile to emphasize three aspects related to the superconducting results. First, given that the electronic Fermi surfaces are concentric-circle like and the phonons show nearly isotropic dispersions along the ΓM and ΓK directions, the adoption of the isotropic Eliashberg equation can be well justified. Secondly, to partially justify the non-SOC calculations, we have compared the EPC strengths of bulk $IrTe_2$ at the Γ point with and without the SOC, showing a negligible difference of less than 1% (Table S6). Moreover, the applicability of the isotropic approximation without the SOC is also well supported by the agreement between the calculated and experimentally observed $T_c$ in bulk $IrTe_2$. In any case, the central physical aspect that the polarity can tune the $T_c$ is expected to stay intact whether the SOC effect is fully accounted for or not. Thirdly, in contrast to the carrier doping mechanism dominated in conventional superconductor/ferroelectric heterostructures [3-5], the mechanism of the ferroelectric-enhanced $T_c$ in the present system is mainly attributed to enhanced interfacial electron-phonon coupling. Here the switching of polarization determines whether the interlayer electrons and interlayer phonons are actively invoked in the EPC process. Such a mechanism bears a resemblance to the basic physics of twisted bilayer graphene, where superconductivity emerges due to interlayer states [48].

Finally, we investigate the ferroelectric switching of band topology in the $IrTe_2/In_2Se_3$ heterobilayers. The different ferroelectric polarizations induce different band alignments and charge transfer between the $IrTe_2$ and $In_2Se_3$ monolayers (see more discussions in the SM [26]), which may play a role in controlling the topological band character of the heterobilayers with inherently strong SOC. Figures 3(a,b) and 3(d,e) show the projected band structures of the $IrTe_2/In_2Se_3$ heterobilayers for both polarizations obtained without and with the SOC. For the downward polarization, although there is no global gap, a local gap exists at every point in the whole Brillouin zone. By introducing a "curved chemical potential" [the blue dashed line in Fig. 3(b)] through the local gap [49], the topological invariant $Z_2$ can be well defined [37], and the band inversions can also be analyzed. Specifically, without the SOC, there exists a small gap of ~ 55 meV between the $In_o+Se_o$ and Te-$p$ bands around the $\Gamma$ point [Fig. 3(a)]. When the SOC is included, the $In_o+Se_o$ and Te-$p$ bands are inverted by crossing the "curved chemical potential" [Fig. 3(b)], which may be accompanied by a topological phase transition. In contrast, for the upward polarization, there is a global gap above the Fermi level, with the energy difference around the $\Gamma$ point as large as 0.93 eV. The inclusion of the SOC can reduce the gap, but cannot close it to induce a band inversion [Fig. 3(e)], resulting in a trivial state. Detailed band evolutions at the $\Gamma$ point are schematically depicted in Fig. S8.

To identify the topological property, we further calculate the topological invariant $Z_2$ using the Wannier charge center (WCC) method [34]. For the downward polarization, an odd number of times of WCC crossing with any arbitrary horizontal reference line is observed [Fig. S9(a)], revealing a topologically nontrivial state ($Z_2$=1). The nontrivial topology is further shown to be rather robust, persisting when subjecting the system to a tensile biaxial strain of 5% (Fig. S10). In addition, the metastable heterobilayer configuration for the downward polarization (stacking 1 in Fig. S1) also shows similar band structures and harbors nontrivial topology (Fig. S10). In contrast, an even number of times of WCC crossing is observed in the upward-polarization case [Fig. S9(b)], leading to its trivial character ($Z_2=0$). As another manifestation of the nontrivial topology in the band structures, the edge states of a semi-infinite slab with Te and Se terminations along the zigzag direction are calculated for both polarizations [Figs. 3(c) and 3(f)]. A pair of topological edge states with Dirac nature at the X point are observed within the bulk band gap for the downward case, while such edge states are absent for the upward case. Here it is noted that, depending on the specific materials combinations of the ferroelectric/superconductor heterostructures, it is expected to realize different band topologies for different functionalities (see more discussions in the SM [26]).

Before closing, we briefly discuss several aspects related to experimental realizations and potential applications of the strong and intriguing predictions made here. First, on the sample preparation aspect, since $In_2Se_3$ monolayers have been widely fabricated [11,50], we suggest to use $In_2Se_3$ as the substrate and transfer or epitaxially grow the $IrTe_2$ monolayer onto the substrate. Our calculations have shown that $In_2Se_3$ can help to stabilize the $IrTe_2$ monolayer (in contrast, the phonon spectra of an $IrTe_2$ monolayer shows significant imaginary frequencies [51]). Secondly, on the aspect of polarization reversal, our calculated energy barrier of an $IrTe_2/In_2Se_3$ heterobilayer against polarization flipping from upward to downward is 0.11 eV, slightly higher than that of a freestanding $In_2Se_3$ monolayer (0.08 eV). The energy difference between the two polarized heterobilayers can help to drive the reversal of the electric polarization, and the small activation barriers can make the kinetic processes accessible at room or relatively low operation temperatures. Thirdly, since materials possessing dual tunabilities in band topology and intrinsic superconductivity are very rare [4,52,53], $IrTe_2/In_2Se_3$ should provide an ideal platform for directly comparing the normal and nontrivial-topology-coupled superconducting state, with minimal chemical or structural disorders.

Finally, the coexistence of superconductivity and topologically nontrivial bands in IrTe$_2$/In$_2$Se$_3$ is highly promising to realize topological superconductivity. The controlled switching of the band topology naturally further serves as a distinctly superior knob in the potential definitive identification of the nature of excitations in the systems, including the Majorana zero modes.

In conclusion, we have demonstrated a simultaneous tuning of $T_c$ and band topology by ferroelectric polarization in the IrTe$_2$/In$_2$Se$_3$ heterobilayers. The opposite polarizations cause different band alignments and charge transfer between the In$_2$Se$_3$ and IrTe$_2$ monolayers when forming the heterobilayers, which play a vital role in enhancing the $T_c$ and inverting the electronic bands. In particular, when the polarization is downward, the charge transfer takes place significantly from the IrTe$_2$ and accumulates at the interface, resulting in stronger electron-phonon coupling that accounts for the higher $T_c$. This downward configuration is also accompanied by the emergence of nontrivial band topology. This study not only showcases a distinct and unprecedented approach to simultaneously tune superconductivity and band topology, but also sheds new light on the underlying mechanisms of ferroelectric-field-induced superconductivity in 2D heterostructures.

**Acknowledgments**

We thank Mr. Leiqiang Li and Dr. Wenjun Ding for helpful discussions. This work was supported by the National Key R&D Program of China (Grant No. 2017YFA0303500), the National Natural Science Foundation of China (Grant Nos. 11634011 and 11974323), the Anhui Initiative in Quantum Information Technologies (Grant No. AHY170000), and the Strategic Priority Research Program of Chinese Academy of Sciences (Grant No. XDB30000000).

**Figures and figure captions**

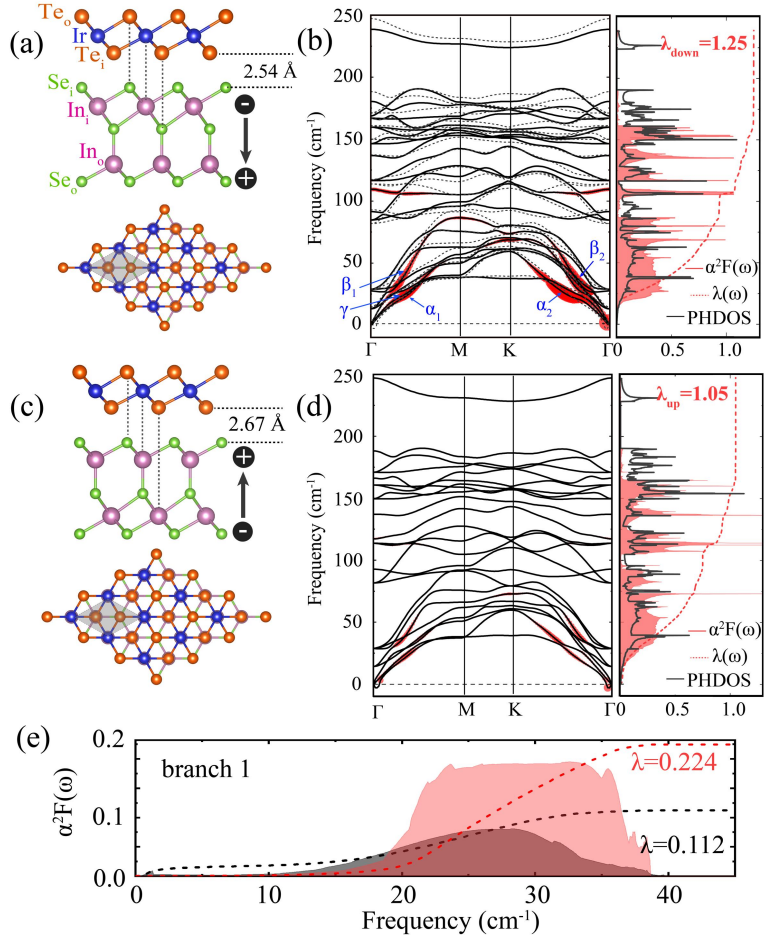

**FIG. 1.** Side and top views of the most stable IrTe$_2$/In$_2$Se$_3$ heterobilayers with (a) downward and (c) upward polarizations. The unit cells are highlighted by the grey shaded areas. (b) Phonon spectra (left panel, solid lines) for the downward polarization, with the branch- and momentum-resolved EPC strengths indicated by the sizes of the red circles. The upward-polarized phonon spectra are superimposed (dashed lines) for comparison. The corresponding phonon density of states (PHDOS), Eliashberg function α$^2$F(ω), and λ(ω) are shown in the right panel. (d) Same as (b) but for the upward polarization. (e) α$^2$F(ω) (shaded region) and λ(ω) (dashed line) for branch 1 for the downward (red) and upward polarizations (dark gray).

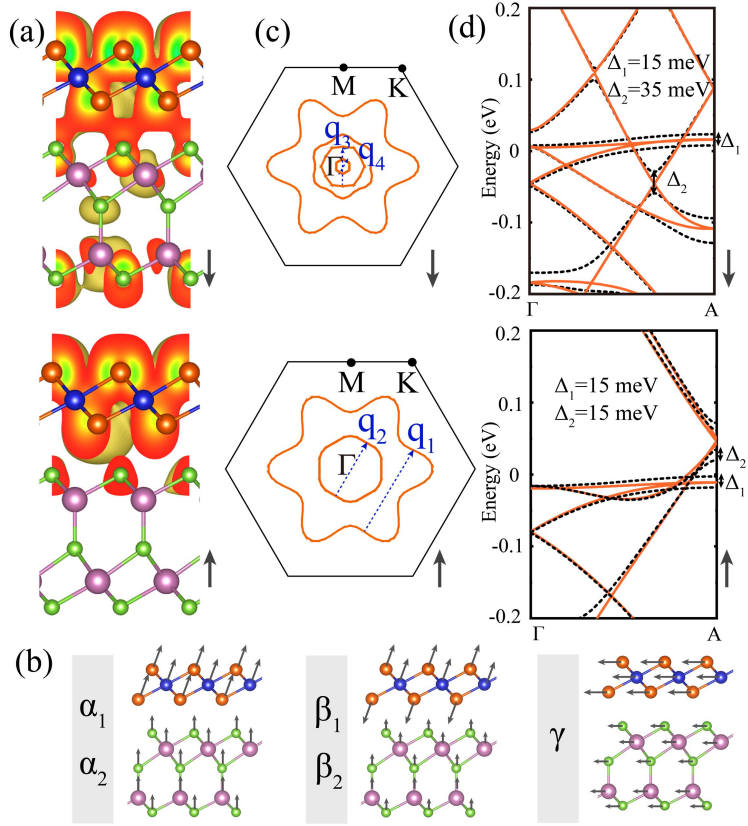

**FIG. 2.** (a) Partial charge distributions within the range of -0.05~0.05 eV for the downward (↓) and upward (↓) polarizations, adopting the same isosurface values. (b) Illustrations of the atomic displacements for the phonon modes $\alpha_1$, $\alpha_2$, $\beta_1$, $\beta_2$, and $\gamma$. (c) Fermi surfaces without the SOC with the nesting vectors $q_1$, $q_2$, $q_3$, and $q_4$ indicated. (d) Band structures of a 6×1×1 supercell before (solid) and after (dashed) the lattice distortion caused by mode $\alpha_1$, with the band splittings indicated.

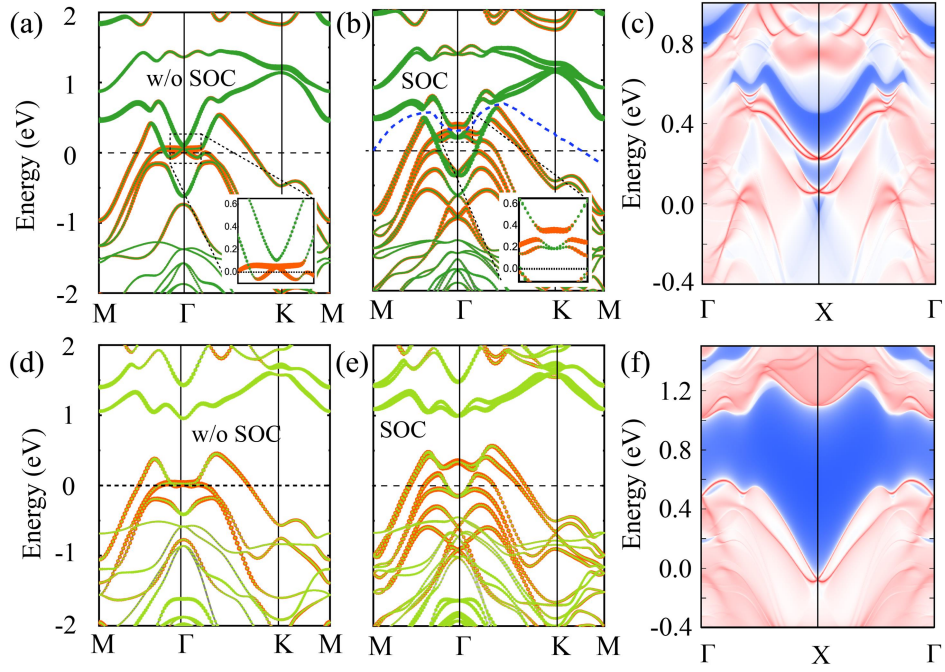

**FIG. 3.** Projected band structures for the downward polarization calculated (a) without and (b) with the SOC. The insets are the zoom-in views. (c) Edge states of the corresponding semi-infinite slab with Te and Se terminations along the zigzag direction. The warmer colors denote higher density of states, and the blue regions denote the bulk band gaps. (d-f) Same as (a-c) but for the upward polarization. In (a, b, d, e), the sizes of the dark green, light green, and orange circles are proportional to the contributions from the $In_o+Se_o$, $In_i+Se_i$, and Te-$p$ orbitals.

**Table and table captions**

**Table 1.** Density of states at the Fermi level $N(\varepsilon_F)$, logarithmic average of the phonon frequencies $\omega_{\log}$, total EPC strength $\lambda$, and superconducting transition temperature $T_c$ of bulk $IrTe_2$ and $IrTe_2/In_2Se_3$ heterobilayers.

| System | $N(\varepsilon_F)$ (Ry$^{-1}$) | $\omega_{\log}$ (K) | $\lambda$ | $T_c$ (K) $\mu^*=0.15\sim0.10$ |
|---|---|---|---|---|
| Bulk $IrTe_2$ | 12.27 | 105.43 | 0.76 | 2.99~4.48 |
| | | | | ~3.0[a] |
| $IrTe_2/In_2Se_3\downarrow$ | 20.88 | 92.33 | 1.25 | 7.70~9.58 |
| $IrTe_2/In_2Se_3\uparrow$ | 19.68 | 88.53 | 1.05 | 5.48~7.14 |

[a]experimental values [15,16].